\documentclass[twocolumn,conference,a4paper]{IEEEtran}

\usepackage{amssymb,amsmath,epsfig,latexsym,graphicx,bm}

\usepackage[left = .56in, top=.75in,right=.56in,bottom=0.8in,nohead,nofoot]{geometry}

\usepackage{epstopdf}
\usepackage{xcolor}
\usepackage{url}
\usepackage{fancyhdr}

\title{Cyberbullying of High School Students in Bangladesh: An Exploratory Study}

\author{\IEEEauthorblockN{\IEEEauthorrefmark{1}
Supriya Sarker, \IEEEauthorrefmark{2}Abdur R. Shahid}
\IEEEauthorblockA{\IEEEauthorrefmark{1}Department of Computer Science and Engineering\\ Chittagong University of Engineering and Technology\\
Chittagong, Bangladesh}
\IEEEauthorblockA{\IEEEauthorrefmark{2}School of Computing and Information Sciences \\Florida International University\\
Miami, Florida, USA\\
\IEEEauthorrefmark{1}sarkersupriya7@gmail.com}
\IEEEauthorrefmark{2}ashah044@fiu.edu}

\begin{document}
\IEEEoverridecommandlockouts

\maketitle
\begin{abstract}
This study explores the cyberbullying experience of the high school students in Bangladesh. The motivation of the work is to identify the internet usage and online activities that may cause cyberbullying victimization of the students of the age between 13 and 18. The study also investigates cyberbullying prevalence and impacts both as victimization and perpetration perspectives, discusses their reporting practices to parents, school officials, other adults and suggest policies to teach cyber safety strategy and generate awareness among students.
\end{abstract}


\section{Introduction}
With the prevalence of smartphones and internet connectivity, the usage of the internet is increasing in Bangladesh. In 2016, 13.2\% individual of the total population of Bangladesh access the internet at home which was 12.1\% in 2015. Users change in one year in the country is 10.4\% \cite{internetlivestats}. Internet usage rates are relatively high among young people. The level of internet usage among young people is seen mostly in the social networking media \cite{sdasia}. As youngsters express many aspects of their lives in the social networking media, their safety in using social media is a growing concern. A particular example of the attempt of blackmail of a 14 years old girl on social media by her uncle and thereby socially embarrassed when she rejected his proposal for an illicit relation (https://www.unicef.org). Nowadays many children in Bangladesh are confronting child abuse through the internet. Cyberbullying, therefore, is one of the major concerns of parents for their children when they use the internet. Therefore, it is necessary to educate youngsters about the potential risks and threats associated with cyberspace and teach themselves how to keep themselves safe. Along with that, it is essential to educate teachers, parents and guardians as they need to help students with this fact \cite{kritzinger2014online}. Despite being a critical problem for students of Bangladesh, cyberbullying in Bangladesh has received little attention not only in the social front but also in the educational research literature. 

\section{Literature Review}
Nowadays cybercrime has been identified as a critical problem in the whole world. Since Bangladesh is somewhat lagging behind in the internet usage compared to other developed countries of the world, cybercrime is still in an emerging stage in the country \cite{kamal2012nature}. The study in \cite{kamal2012nature} found that although cybercrime has not received enough attention in the field of research in Bangladesh, many people are becoming victims through the internet. In the recent years, the rate of active users of the internet are growing rapidly. According to Google, the number of active internet users has stood at 4 crores with 35\% using it every day and the number will be 9 crores within the year 2020 (published on Dec 1, 2017) \cite{daily.star}. According to Socialbakers statistics, 73\% of Facebook users are in the age group of 13-25 years. This image reveals that adolescents and teenagers are ahead in using Facebook and other social media \cite{daily.asian.age}. Cyberbullying becomes a common phenomenon of online violence on social media. A study conducted in 2012 has revealed that at least 800,000 young people have become the victim of online violence on Facebook. The number of middle school children who experienced cyberbullying was almost double. 49.5\% of students reported being a victim of online bullying. This leads to mental damage, even while more vulnerable, the students move towards suicide \cite{daily.observer}. \\Global research reflects that cyberbullying is being proved as a universal invasion that concerns the middle primary and secondary school students. i-Safe (http://isafe.org/) has reported on a survey of 1500 students age range of 9 to 13 years old where 42\% students reported that they have been bullied online \cite{smith2006investigation}. More research works have been published on cyberbullying of school students in the United States. Middle school students of grade 6, 7, and 8 in the southeastern and northwestern United States have surveyed with a set of questionnaires consists of the Olweus Bully/Victim Questionnaire and 23 standard questions \cite{popovic2011prevalence}.
The research work in \cite{smith2006investigation} reported a study on cyberbullying conducted with the students from grade 6 to 9 in five schools in British Columbia, Canada. The study has quantified digital technology such as computer and cell phone usage, identified the type, extent and impact of cyberbullying incidents both as the victim and perpetrator perspective and investigated the online behaviour of young students. The INQUIRER (www.theinquirer.net) has published a statistics studied by Queensland University of Technology. It reported that 13\% of students had experienced cyberbullying at the age of 8 and that of 25\% of participants reported that they had known someone who experienced cyberbullying. Moreover, half of the students thought that the phenomenon has been rising \cite{smith2006investigation}. 
In Serbia 387 middle school adolescences have surveyed to investigate the prevalence of cyberbullying. The results showed that most of the students regularly used the internet and all of them used the cell phone. 10\% reported to bully others online with 20\% of them was the victim of cyberbullying \cite{popovic2011prevalence}. 
A study carried out during May 2010 to explore the cyberbullying experience of Turkish students which has revealed that male students were more involved in cyberbullying compared to female students and the majority of the participant students have not aware of any cyber safety strategy \cite{yilmaz2011cyberbullying}. \\
There is little research on cyberbullying in Bangladesh that published in a few journals and to the best of our knowledge, there is no research done on the online bullying of middle school students in Bangladesh. It proves that research on the crucial matter is still in an elementary stage. The study in \cite{monni2016investigating} explored the causes of cyberbullying and its sociopsychological impacts on girls in Bangladesh. Our goal is to promote cyber-bullying research and to determine the present condition and impact of cyberbullying among students, especially high school students. The way cyberbullying research can be carried forward is by surveying the current state of online activities and cyber violence status and impacts on Bangladeshi high school students.

\section{Problem Statement, Research Questions and Methodology}
\begin{table}
\centering
\caption{Age of the Respondent}\label{Table:1}
  \begin{tabular}{|c|c|c|}
\hline
\bf{Age}&\bf{Frequency}&\bf{Percentage} \\\hline
13-14 & 3 & 14.29\% \\ \hline
15-16 & 16 & 76.19\% \\ \hline
17-18 & 2 & 9.52\% \\ \hline
  \end{tabular}
  
\end{table}

This work reports on an investigation of current online involvement among high school learners in Bangladesh that victimizing the high school students. To develop the cyber safety awareness among high school learners in Bangladesh, it is important to gather and analyze statistical information on the cyber activities done by them. 
\subsection{Problem Statement}
Our research focuses primarily on the high school learners in Bangladesh between the ages of 13 and 18 to obtain statistical data regarding cyberbullying. 
\subsection{Research Questions}
Our motivation is to answer to the following research questions: 
\begin{itemize}
  \item RQ1: Which online activities are victimizing high school students in Bangladesh? 
  \item RQ2: What initiatives can be taken to improve online safety among high school learners?
\end{itemize}
\subsection{Methodology}
The data for this report is being collected through a survey among high school learners of different areas of Bangladesh. We have studied the various structure of survey questionnaire and selected the most applicable questions for middle school students of Bangladesh mentioned in \cite{smith2006investigation}, \cite{li2010cyberbullying}, \cite{wired.safety}. Participants are provided with 26 multiple choice questions. As no identifying data is being collected, it is assured that responses are anonymous. The structure of the questionnaire includes four domains: 1) Demographic information of the participants, 2) Cyberbullying experience of the participants as a perpetrator and as a victim, 3) witness of cyberbullying incidents, 4) opinion and suggestions regarding cyberbullying. The questionnaire provides a definition of cyberbullying, with examples, to make sure a general understanding of the phenomena. Data on cyberbullying is being collected during May 2018 from 21 students (12 male and 9 female) of class 8, 9 and 10, representing both urban (38\% of participants) and rural(62\% of participants) areas of Bangladesh. Table \ref{Table:1} shows the age range and frequency of the respondents.

\section{Online Activities Engaged by high school students}    
The survey investigates the length of time the high school learners spent using cell phones and on the internet. 18 out of 21 respondents (86\%) indicate that they use the internet at home and 3 out of 21 respondents (14\%) indicate that they do not. The students of class 9 and 10 have the tendency to spend a longer time than that of class 8. 
\subsection{Cell phone usage at home and school}  
As the use of cell phone and internet is increasing among students, social networking websites are becoming very popular for social and personal communication. To the youngsters, online media are so open that their personal information reveals very easily to others \cite{monni2016investigating}. According to our best knowledge, there is no high school in Bangladesh where the cell phone is allowed for learners. 67\% learners do not use the cell phone in school hours while few learners use the cell phone at school regularly (24\%) and at times (9\%), respectively.
\subsection{Most used devices}  
\begin{figure}
\centering
\footnotesize
\renewcommand{\arraystretch}{2}
\includegraphics[width=0.66\linewidth]{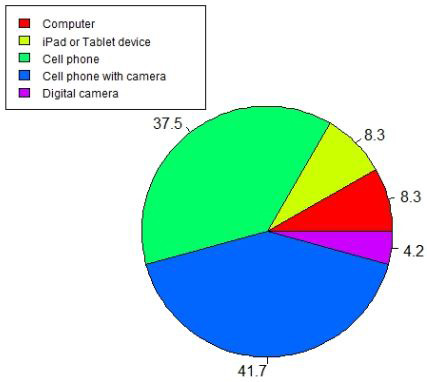}
\caption{Most used devices by high school learners.} 
\label{Fig:1}
\end{figure}

The most used electronic devices by high school learners are the cell phone with picture taking capabilities (41.7\%) and cell phone with internet capabilities (37.5\%). Only a few learners use the computer with email (8.3\%), iPad or Tablet device with internet capabilities (8.3\%), and digital camera (4.2\%). Figure \ref{Fig:1} shows the most used devices by high school learners. In case of cyberbullying, personal information, images are captured and videos of victims are recorded through digital devices and posted in the internet \cite{campbell2005cyber}. This trend is increasing among students with the wide usage of digital devices.

\subsection{Most favorite online activities}
Their most favourite online activities include communicating with friends (48\%) and surfing to look for stuff or learn new things (52\%). Other online activities such as playing online games (29\%), using social networking (Facebook, Twitter, Instagram, Pinterest) (24\%), doing homework (5\%) has been noted. 

\section{Results}
The survey investigates the reaction of school learners to the phenomena of cyberbullying and finds that more than half of the learners (52\%) report it as upsetting and 24\% respondents mark it as very upsetting. When young learners are being asked how they feel about cyberbullying, 48\% report that cyberbullying is a serious problem and we need to stop it, 43\% report that cyberbullying is too bad but we cannot do anything. The next section will investigate cyberbullying more detail. 
\begin{figure}
\centering
\footnotesize
\renewcommand{\arraystretch}{2}
\includegraphics[width=0.66\linewidth]{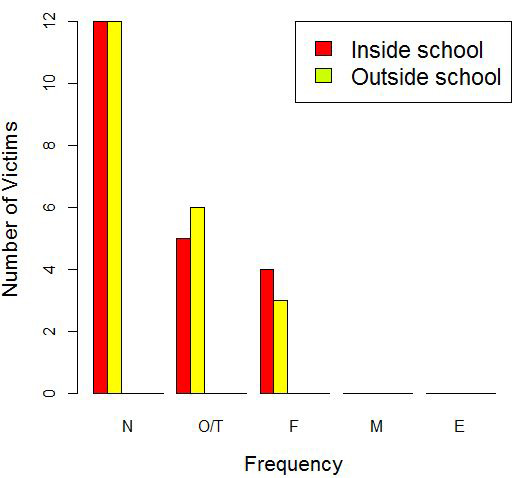}
\caption{Cyberbullying Victimization inside and outside of the school} 
\label{fig:2}
\end{figure} 
\subsection{Cyberbullying experiences as victim}
Of the learners of Class 8, 9 and 10, 57\% of the respondents report that they have never been cyberbullied outside of the school campus. Of the victims, 24\% report that they have been cyberbullied ‘once/twice’ and 19\% of the learners have cyberbullied ‘a few times’ outside of the school. 57\% of respondents report that they have never been cyberbullied in the school campus, 28.57\% of them have cyberbullied ‘once/twice’ and 14.29\% of them have cyberbullied ‘a few times’. Figure \ref{fig:2} shows the cyberbullying victimization in school and outside of school where N, O, T, F, M, E represents Never, Once, Twice, Few, Many, Everyday, respectively. 

\subsection{Cyberbullying experiences as perpetrator}
\begin{figure}
\centering
\footnotesize
\renewcommand{\arraystretch}{2}
\includegraphics[width=0.66\linewidth]{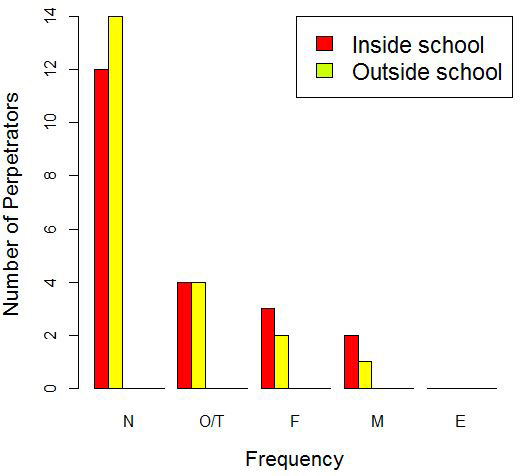}
\caption{Cyberbullying Perpetration inside and outside of the school } 
\label{fig:3}
\end{figure}

While 57.14\% respondents state that they have never been cyberbullied others outside of the school, 19.04\% of respondents report that they have cyberbullied others outside of the school once or twice. When they are being asked how often they have cyberbullied others inside of school campus they note that 19\% respondents have done that ‘once/twice’ and 14\% respondents have done that ‘a few times’ and 9.5\% have done that ‘many times’. The higher rate of cyberbullying has done by male compared to female students. Figure \ref{fig:3} shows the cyberbullying perpetration in school and outside of school where N, O, T, F, M, E represents Never, Once, Twice, Few, Many, Everyday, respectively. 

\subsection{What to do while experiencing cyberbullying?}
Learners are being asked what have they done after being cyberbullied. Most of the victims (43\%) report that they would tell a friend, 33\% of them note that they would tell the cyberbully to stop, 24\% of the respondents mark that they would log off from online or get away from the cyberbully. Only 5\% of students report that they would tell a teacher or school staff. This proves that they are not willing to share this kind of incidents with any member of the school. Rather than that, they are likely to share cyberbullying incidents with their friend(s). The survey finds that 28\% of victims think that telling someone about cyberbullying incidents to improve the situation, 14\% victims of cyberbullying never tell anyone, 9.5\% victims think nothing changes after telling someone and 5\% note that it gets worse after telling someone. 
    
\subsection{Who come up with help?}  
\begin{figure}
\centering
\footnotesize
\renewcommand{\arraystretch}{2}
\includegraphics[width=0.66\linewidth]{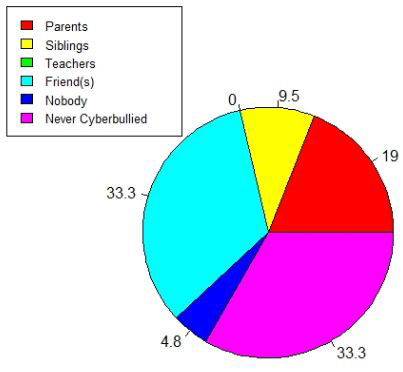}
\caption{If cyberbullied, who tried to help?} 
\label{fig:4}
\end{figure}

Students are being asked who tried to help when they have been cyberbullied. 33.3\% victims got help from their friend(s), 9.52\% of victims got help from siblings, and only 14.29\% victim got help from their parents, and 4.76\% victims got help from nobody. No student report getting help from the teacher or school staff. Figure \ref{fig:4} shows from whom students get help.
\subsection{Remarks or comments made by cyberbullies}  
 Most of the comments are made on self-worth (24\%) and appearance (14\%). 5\% of respondents report that they face hurtful comments regarding their intelligence. 9.5\% students note that they were asked to introduce their female friend(s) for a purpose of making romantic relation to a person whom they met online. There are lot of examples of broken relationship results in cyber harassment of girls \cite{monni2016investigating}.
\subsection{Assumed reasons behind cyberbullying}  
Young learners mark the reasons behind cyberbullying other people. 38\% of them note that cyberbullying is done as fun, 29\% of respondents report that cyberbullies are mean, 24\% report that cyberbullies generally do this as a defensive mechanism, 14\% report cyberbullies to have family problems, 5\% report cyberbullies have taken cyberbullying as a cool fact, they are jealous and angry. 10\% of learners specify cyberbullying as the influence of bad company. 
\subsection{Cyberbullying bystanders}    
 When they are asked about their normal responses being a witness of cyberbullying incidents, 62\% learners mark that they would report the incidents to someone who can help the victim, 19\% learners note that they would object to others, but not directly to the cyberbullies, 14\% report that they would try to help or befriend the victim privately, 10\% of learners would object to cyberbully, 4.7\% report that they would watch but would not participate. 
\subsection{Reasons of unwillingness to discuss} 
24\% of students notes that they could get themselves into trouble, even if they had done nothing wrong, 19\% of the respondents mark that school staffs or teachers would not believe or do anything to stop cyberbullying. The other reasons are students would make fun of them at school (14.29\%), cyberbully could get back and make things even worse (19.04\%), the learners fear that their parents could find out their online activities and restrict the facilities (9.52\%). Among the respondents, 14.3\% of the students note that they believe that they must learn to stop cyberbullying and help themselves. 

\section{Discussion}
\subsection{Opinion of the Students}
They are asked how cyberbullying affect high school learners. The learners report that it could affect a student mentally (47.62\%), create apathy in going to school (38.09\%), could create fear to use the internet (57\%). Besides, it could harm to the physical condition of the students, 9.52\% of the learners report it could cause reluctance to make friends.
\subsection{Status of cyber safety strategy of high school students}
Above 71.43\% of students indicate that they have not aware of any cyber safety strategy. This reflects a huge urgency of generating cyber safety awareness among young online users. Few of them are aware of some cyber safety strategies (28.57\%) who have learned cyber safely strategies from their friends (19.05\%), siblings (5\%) and teachers and school staffs (4\%). It implied that teachers and schools administration need to be more concerned about cyberbullying both outsides of the school campus and in the school. Almost 90.48\% learners note that teaching cyberbullying in school curriculum would help them to learn about it.    
\subsection{Suggestions to develop cyber safety strategy}
School can play a vital role to teach cyber safety by developing their own ICT policy, adding cyberbullying in the curriculum, regular discussion and experience sharing with students, teaching aids like posters, pamphlets, pledge, wallet booklets (both in English and Bengali language) can be used. These programs should include parents to teach how to respond to cyber harassment of their children \cite{sonhera2012proposed}. Nationwide cybersecurity awareness initiatives can be taken to increase awareness \cite{mohammed2016model}. 
\subsection{Limitations of the study}
The students in rural areas are less familiar with the related terms of online harassment than the students of urban areas. The students who are unaware of it, may not put appropriate answers to some of the cyberbullying questions. As cyberbullying is a shameful incident some students may be unwilling to share their cyberbullying experiences in the survey. There is a lack of resources and publications on cyberbullying of high school students in Bangladesh perspectives. Therefore, publication references on Bangladesh for the study may be insufficient. 

\section{Conclusions}
A quantitative approach is used in the study to gather data regarding cyberbullying of adolescences and teen students in Bangladesh by means of a survey that focuses on identifying the online engagement and cyber safety awareness status of high school students. The findings have been discussed and by considering the opinions of the students, some suggestions have been proposed to educate young learners and generate cyberbullying awareness among learners, parents and educators.





\begin{thebibliography}{10}

\bibitem{internetlivestats}
(2016) {I}nternet {L}ive {S}tats. [{O}nline]. {A}vailable:.
\newblock \url{http://www.internetlivestats.com/internet-users/bangladesh/}.
\newblock Accessed: 2018-06-24.

\bibitem{wired.safety}
(2016) {W}ired {S}afety website. [{O}nline]. {A}vailable:.
\newblock \url{https://www.surveymonkey.com/r/StudentCyberbullyingSurvey}.
\newblock Accessed: 2018-06-24.

\bibitem{daily.asian.age}
(2017) {D}aily {A}sian {A}ge. [{O}nline]. {A}vailable:.
\newblock
  \url{https://dailyasianage.com/news/46958/social-media-trends-usages-in-bangladesh}.
\newblock Accessed: 2018-06-24.

\bibitem{sdasia}
(2017) {SD ASIA} [{O}nline].{A}vailable:.
\newblock \url{https://sdasia.co/2016/02/10/cyber-bullying-in-bangladesh/}.
\newblock Accessed: 2018-06-24.

\bibitem{daily.star}
(2017) {T}he {D}aily {S}tar. [{O}nline]. {A}vailable:.
\newblock
  \url{https://www.thedailystar.net/business/4cr-internet-users-bangladesh-google-1498903}.
\newblock Accessed: 2018-06-24.

\bibitem{daily.observer}
(2018) {T}he {D}aily {O}bserver. [{O}nline]. {A}vailable:.
\newblock \url{http://www.observerbd.com/details.php?id=119829}.
\newblock Accessed: 2018-06-24.

\bibitem{campbell2005cyber}
Marilyn~A Campbell.
\newblock Cyber bullying: An old problem in a new guise?
\newblock {\em Journal of Psychologists and Counsellors in Schools},
  15(1):68--76, 2005.

\bibitem{kamal2012nature}
M.~M. Kamal, I.~A. Chowdhury, N.~Haque, M.~I. Chowdhury, and M.~N. Islam.
\newblock Nature of cyber crime and its impacts on young people: A case from
  bangladesh.
\newblock {\em Asian Social Science}, 8(15):171, 2012.

\bibitem{kritzinger2014online}
E.~Kritzinger.
\newblock Online safety in south africa-a cause for growing concern.
\newblock In {\em Information Security for South Africa (ISSA), 2014}, pages
  1--7. IEEE, 2014.

\bibitem{li2010cyberbullying}
Q.~Li.
\newblock Cyberbullying in high schools: A study of students' behaviors and
  beliefs about this new phenomenon.
\newblock {\em Journal of Aggression, Maltreatment \& Trauma}, 19(4):372--392,
  2010.

\bibitem{mohammed2016model}
S.~Mohammed and E.~Apeh.
\newblock A model for social engineering awareness program for schools.
\newblock In {\em Software, Knowledge, Information Management \& Applications
  (SKIMA), 2016 10th International Conference on}, pages 392--397. IEEE, 2016.

\bibitem{monni2016investigating}
S.~S. Monni and A.~Sultana.
\newblock Investigating cyber bullying: Pervasiveness, causes and
  socio-psychological impact on adolescent girls.
\newblock {\em Journal of Public Administration and Governance}, 6(4):12--37,
  2016.

\bibitem{popovic2011prevalence}
B.~Popovi{\'c}-{\'C}iti{\'c}, S.~Djuri{\'c}, and V.~Cvetkovi{\'c}.
\newblock The prevalence of cyberbullying among adolescents: A case study of
  middle schools in serbia.
\newblock {\em School psychology international}, 32(4):412--424, 2011.

\bibitem{smith2006investigation}
P.~K. Smith, J.~Mahdavi, M.~Carvalho, and N.~Tippett.
\newblock An investigation into cyberbullying, its forms, awareness and impact,
  and the relationship between age and gender in cyberbullying.
\newblock {\em Research Brief No. RBX03-06. London: DfES}, 2006.

\bibitem{sonhera2012proposed}
N.~Sonhera, E.~Kritzinger, and M.~Loock.
\newblock A proposed cyber threat incident handling framework for schools in
  south africa.
\newblock In {\em Proceedings of the South African Institute for Computer
  Scientists and Information Technologists Conference}, pages 374--383. ACM,
  2012.

\bibitem{yilmaz2011cyberbullying}
H.~Yilmaz.
\newblock Cyberbullying in turkish middle schools: An exploratory study.
\newblock {\em School Psychology International}, 32(6):645--654, 2011.

\end{thebibliography}
\end{document}